\documentclass[aps,prl,showpacs,twocolumn]{revtex4}
\usepackage{amssymb,amsmath, psfrag, epsfig}

\newcommand{\bel}[1]{\begin{equation}\label{#1}}

\newcommand{\be}{\begin{equation}}
\newcommand{\ee}{\end{equation}}
\newcommand{\eeq}{\end{equation}}
\newcommand{\baS}{\begin{eqnarray}}
\newcommand{\ba}{\begin{eqnarray}}
\newcommand{\ea}{\end{eqnarray}}
\newcommand{\rf}[1]{(\ref{#1})}

\newcommand{\bi}{\bibitem}

\newcommand{\tn}{$\theta$-neuron}

\newcommand{\qif}{quadratic integrate and fire model}

\def\p{\partial} 
\def\xt{x_{th}}

\begin{document}

\title{Transient termination of  synaptically sustained spiking by stochastic inputs\\  
in a pair of coupled Type 1 neurons}

\author{Transient termination of  synaptically sustained spiking by stochastic inputs\\  
in a pair of coupled Type 1 neurons}
\email{boris.gutkin@ens.fr}
\affiliation{Group for Neural Theory, 
D\'epartment des Etudes Cognitives, Ecole Normale Sup\'erieure, 
5, rue d'Ulm, 75005 Paris, France}
\author{J\"urgen Jost}
\altaffiliation{Institute des Hautes Etudes Scientifiques, 
35, route de Chartres, 91440 Bures-sur-Yvette, France}
\affiliation{Max Planck Institute for Mathematics in the Sciences, 
Inselstr. 22,  04103 Leipzig, Germany}
\altaffiliation{Institute des Hautes Etudes Scientifiques, 
35, route de Chartres, 91440 Bures-sur-Yvette, France}
\author{Henry C. Tuckwell} 
\email{jost, tuckwell@mis.mpg.de} 
\affiliation{Max Planck Institute for Mathematics in the Sciences, 
Inselstr. 22,  04103 Leipzig, Germany}

 \begin{abstract}  We examine the  effects of stochastic input currents on the firing behavior of two 
 excitable  neurons coupled with fast
excitatory synapses. In such cells
(models), typified by the quadratic integrate and fire model, mutual synaptic
coupling can cause sustained firing or oscillatory behavior which is necessarily
antiphase. Additive Gaussian white noise can transiently 
terminate the oscillations, hence destroying the stable limit
cycle. Further application of the noise may return the system to
spiking activity. In a particular noise range, the transition times
between the oscillating and the resting state are strongly
asymmetric. We numerically investigate an approximate
basin of attraction, $\cal{A}$, of the periodic orbit and use Markov process theory to
explain the
firing behavior in terms of  the probability
of escape of trajectories from $\cal{A}$. 
\end{abstract}
\maketitle

Recent experiments \cite{Ta1}  have shown that the spiking patterns of regular spiking and fast spiking 
 neurons in the rat somatosensory cortex
exhibit  Type 1 and Type 2 behavior, respectively. Such differences
were originally found  by
Hodgkin \cite{Ho}  in his investigations of the responses of
squid axon preparations  to applied currents.  In some cases, the
frequency of firing rose smoothly from zero as the current increased
whereas in others, a train of spikes with a non-zero minimal frequency suddenly occurred at a particular
input current. Cells that responded in the first manner were called Class 1 (which we call Type 1)
whereas cells with discontinuous
frequency-current curves were called Class 2 (Type 2).  Mathematical explanations for the two types are
found in  the bifurcation  which accompanies the transition
from rest state to the periodic firing mode. For Type 1 behavior, a resting potential
vanishes via a saddle-node bifurcation whereas for Type 2 the instability
of the rest point is due to an Andronov-Hopf bifurcation; see for
example \cite{Ri}.

Here we analyze the effects of stochastic inputs on the firing behavior of
coupled  neurons of Type 1.    Although there have been many  studies on single
neurons of this type, \cite{Gu1,Br,Li}, the
effect of noise on systems of coupled Type 1 neurons has not been
extensively 
investigated \cite{Gu3,Ca}.
We identify a novel effect of noise on the firing sustained by recurrent excitatory synapses
in a pair of Type I neurons: weak noise effectively terminates the firing by destroying the stable limit cycle. 
Stronger noise can lead to intermittent oscillatory behavior. Particularly unexpected are simulations that 
suggest that such an effect is generic and does not depend on the noise model, although 
our  focus is on Gaussian white noise.   We explore two  analytical approaches
for explaining the "inhibitory" effect of the noise, one via first-exit time theory and the other
using moment differential equations. 

{\it The quadratic integrate and fire model and the \tn.} --- 
Computational models which include details of the complex anatomy and physiology of cortical neurons are too
complicated to analyze mathematically. 
However, we can take advantage of the generic nature (as the local
normal form of a saddle node bifurcation) of a relatively
simple  neural model that exhibits Type 1 firing behavior. This is  
 the quadratic integrate and fire (QIF) model \cite{La} for which 
 \vspace*{-2ex}
\bel{0}
\dot x=(x-x_R)^2 + \beta,
\vspace*{-2ex}
\ee 
 where $x$ is interpreted as the membrane potential of the neuron, 
$x_R$ is its resting value and $\beta$ is the mean 
input. Once the value $x_C$ of $x$ is so large that the r.h.s. of (1) is positive, it will become infinite 
in a finite time (of the order of $1/x_C$) and then has to be reset to
$-\infty$. The upward excursion and resetting constitute   a ``spike''
in this model. Problems with infinite values can be
 avoided by  applying
the  transformation 
$x-x_R=\tan\frac{\theta}{2}$,
where $\theta$ takes values in $[0,2\pi]$, i.e., on the unit
circle $S^1$ 
when we identify $0$ and $2\pi$. This yields the \tn\ model \cite{Er,Gu1}
\vspace*{-2ex}
\bel{tn2}
\dot \theta=1-\cos\theta+(1+\cos\theta)\beta, 
\vspace*{-2ex}
\end{equation}
where $\theta=\pi$ corresponds to a spike of the neuron.  
 Both of these equivalent formulations have been used previously for simulation and analysis of neural dynamics \cite{Gu1,La}.

 We consider the case of  two coupled
identical QIF neurons $i=1,2$ with noise terms
as follows \cite{Gu3} 
\vspace*{-2ex}
\begin{eqnarray}  
\label{1}
dX_1&=&[(X_1-x_R)^2 + \beta + g_{s}X_3]dt + \sigma dW_1\\
\label{2} 
dX_2&=&[(X_2-x_R)^2 + \beta + g_{s}X_4]dt + \sigma dW_2 \\
\label{3}
dX_3&=& \big[-\frac{X_3}{\tau} + F(X_2)\big]dt   \\
\label{4}
dX_4&=&\big[-\frac{X_4}{\tau} + F(X_1)\big]dt 
\vspace*{-2ex}   
\end{eqnarray}
where  $X_1, X_2$ are random
processes corresponding to the membrane potentials of the neurons
while $X_3$ ($X_4$) is the synaptic input from neuron 2(1) to neuron 1(2).  In these equations, $g_s$ is the
coupling strength between the neurons and  
$W_1$ and $W_2$ are independent standard Wiener
processes which enter with amplitude $\sigma$. 
The noise terms  represent fluctuations in nonspecific inputs to each
neuron as well as possibly intrinsic membrane and channel noise.  The
function $F$ is given by $F(x)= 1 + \tanh (x-x_{th})$,
where $x_{th}$ characterizes the threshold effect of synaptic
activation \cite{Tu2}, so the variables $X_3, X_4$   take  values in
the interval $[0,2]$. 
The corresponding \tn\ equations  then are
\bel{tn2}
\vspace*{-1ex}
\begin{split}
d \Theta_i=&(1-\cos\Theta_i-  (\sigma^2/2) \sin\Theta_i(1+\cos\Theta_i)[\beta +
g_{s}X_{i+2}])dt\\ 
&+ \sigma (1+\cos\Theta_i)dW_i\quad (i=1,2),
\end{split}
\vspace*{-2ex}
\end{equation}
where $\Theta_i=\pi$ corresponds to a spike of neuron $i$. Note
that at this spike point, the effect of the noise term vanishes. As
verified by Gutkin (unpublished), this, together with the strictly
positive contribution of the term $1-\cos\Theta_i$, ensures that
the spike point $\Theta_i=\pi$ can only be passed in the direction
of increasing values of $\Theta_i$. Therefore, this model is
equivalent to the \qif\ with resetting at $\infty$.

For the purpose of this report we choose a negative value for $\beta$ so that  each  neuron in
isolation will not fire by itself when its potential is near the
resting value $x_R$, but only when  perturbed  beyond a threshold $x_T$.  For the model described by  (3)-(6), without noise,    inducing
firing in one neuron by perturbing it beyond threshold leads to sustained firing in both neurons when the 
coupling strength is above  the bifurcation value 
$g_s=g_s^\ast$. At that value, two heteroclinic orbits
between the unstable rest points where one of the neurons is at $x_R$, the
other at $x_T$, turn into a periodic orbit of antiphase
oscillations.  We then have two stable attractors, the stable rest
point where both neurons take the value $x_R$, and the antiphase
oscillator.  We note that the dynamics are equivalent for both versions of the model circuit: the QIF and the $\theta$-neuron.

\begin{figure}[!t]
\begin{center}
\centerline\leavevmode\epsfig{file=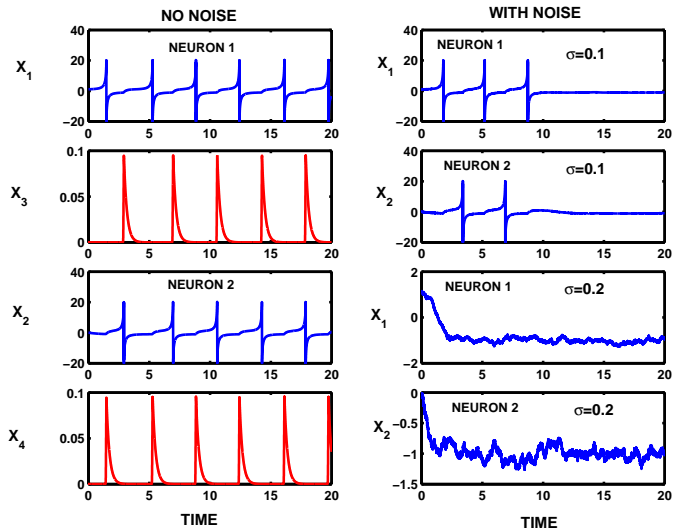,width=3.5in}
\end{center}
\caption{On the left are shown the solutions of \rf{1}-\rf{4} for two coupled
QIF model neurons with the standard parameters and without noise.
$X_1$ and $X_2$ are the potential variables of neurons 1 and 2 and $X_3$ and
$X_4$ are the inputs to neurons 1 and 2, respectively. On the right are shown
examples of trajectories of the potential variables when there is noise, $\sigma = 0.1$,
in the top two parts and $\sigma=0.2$ in the lower two parts. Note the absence
of spikes in the trial for the larger noise case.}
\label{fig:wedge}
\end{figure}

{\it Results and theory.}--- In the numerical work, the following constants are employed as the standard set 
throughout. $x_R=0$,  $x_{th} =10$,  
$\beta =-1$, $g_{s}=100$ and  $\tau=0.25$. In our simulations, we reset the state variables $X_1,X_2$ to the value
$-x_C$ when they reach or exceed the value
$x_C=20$. 
 The initial values of the
neural potentials are $X_1(0)=1.1$, $X_2(0)=0$ and the initial values of the synaptic variables are $X_3(0)=X_4(0)=0$. 
Results such as those in Figure 1 are obtained. The spike trains of the two
coupled neurons and their synaptic inputs are shown on the left for no noise.  The firing settles down to be
quite regular and the periodic orbit  is in part shown in the $(x_1,x_2)$-plane as the red curve in Figure 2.  
 
   \begin{figure}[!t]
\begin{center}
\centerline\leavevmode\epsfig{file=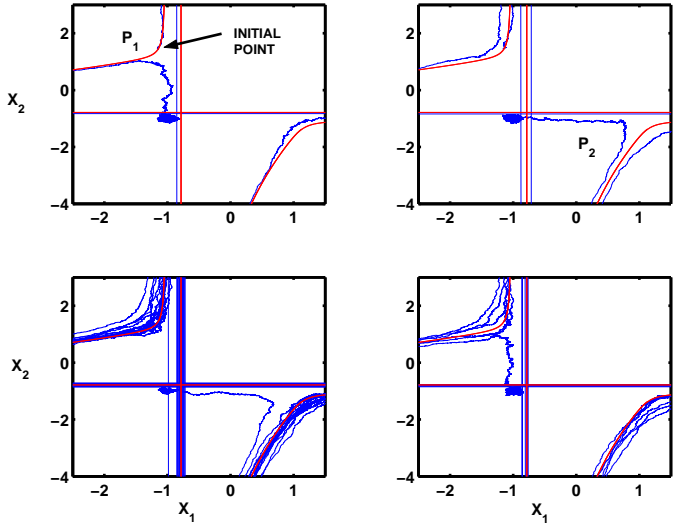,width=3.5in}
\end{center}
\caption{Four examples which illustrate the noise-induced collapse away (blue paths) from the basin of attraction of 
periodic orbit, denoted by the red curve,  in two coupled QIF neurons. The parameters are the standard set, with
$\sigma=0.1$, and 
the initial point is the same in all cases, being on the periodic orbit as indicated top left. 
The departure points are located either near $P_1$ or $P_2$. In the top left example there is
a spike in neuron 2, then a spike in neuron 1 (both clipped) followed by escape from the
basin of attraction before a second spike can occur in neuron 2. }
\label{fig:wedge}
\end{figure}

     \begin{figure}[!t]
\begin{center}
\centerline\leavevmode\epsfig{file=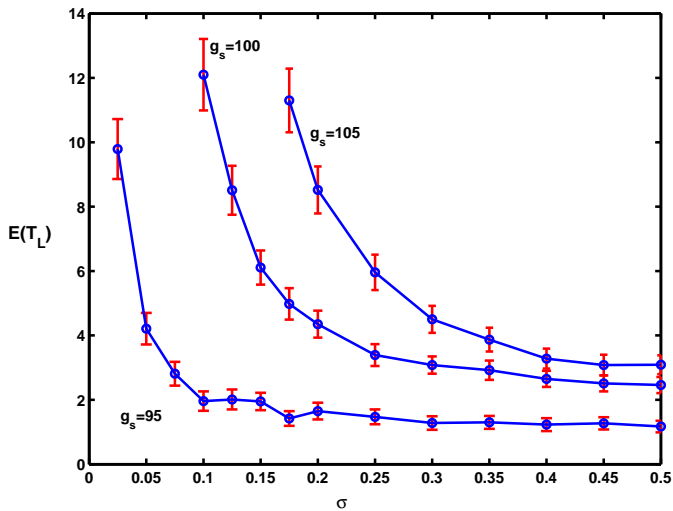,width=3.5in}
\end{center}
\caption{The mean of the time $T_L$ to the last spike in neuron 1 versus the noise amplitude.
This quantity measures the exit time from the basin of attraction of the periodic orbit.
The 4-dimensional process $(X_1,X_2,X_3,X_4)$ is initially on the periodic orbit. Error bars
denote 95\% confidence intervals (500 trials). }
\label{fig:wedge}
\end{figure}
The effects of  weak noise on the spike trains are shown in the right column of Figure 1.  In the
top portion an example of the trajectory for $\sigma=0.1$ is shown. Here three spikes arise in neuron 1
and two in neuron 2, but the time between spikes increases and eventually the orbit collapses away
from the periodic orbit. In the example (lower part) for $\sigma=0.2$ there are no spikes in either neuron.
Extensive simulations showed that orbits tended to collapse away from
the periodic orbit in the vicinity of the points $P_1$
and $P_2$, as shown in four trials in Figure 2. Here, the red curve depicts
the  stable periodic orbit in the absence of noise and the blue curves
 are trajectories with noise, $\sigma=0.1$. These random paths
always start precisely at the same point in $R^4$ on the periodic orbit
- marked as  ``initial point''  .   The
number of orbits completed is a random variable which we may quantify
using $T_L$, which is the time of occurrence of the last spike in
neuron 1.  
Inspection of histograms of $T_L$-values, based on 500 trials, for $\sigma=0.1$ and $0.45$,
shows  that the spiking stops after approximately an integer
multiple of the first spike time. Furthermore, as $\sigma$ increases,
the number of occasions on which no spike was generated
increases. This is further illustrated in Figure 3 where the mean of
$T_L$ is plotted against $\sigma$. Hence it is clear that noise tends to curtail firing. The theoretical basis of these plots 
is outlined in the next section. 
{\it Exit-time and orbit stability.}---
If a basin of attraction for the periodic orbit can be found, then the probability that the process with noise
escapes from  this basin gives the probability, in the present context,
that spiking will cease. Since the system (3)-(6) is Markovian, we may apply standard first-exit time theory \cite{Tu1}. 
Let $A$ be a set in $(S^1)^2 \times [0,2]^2$ and let $y_1, y_2$
be the values assumed by $\Theta_1$ and $\Theta_2$ and let $y_3, y_4$ be the
values assumed by the synaptic input variables $X_3$ and $X_4$. 
The probability $p(y_1,y_2,y_3,y_4)$ that the process ever escapes from
$A$ is given by
\vspace*{-2ex}
\begin{eqnarray}
     \label{fk1}
 {\cal L}p &\equiv& \frac{\sigma^2}{2}(1+\cos y_1)^2 \frac
 {\p^2p}{\p y_1^2}+
 \frac{\sigma^2}{2}(1+\cos y_2)^2 \frac {\p^2p}{\p y_2^2}      \\
\nonumber
 &+& [1-\cos y_1-(\sigma^2/2)\sin y_1(1+\cos y_1)(\beta +
g_{s}y_{3})]\frac{\p p}{\p y_1} \\
\nonumber
 &+&
 [1-\cos y_2-(\sigma^2/2)\sin y_2(1+\cos y_2)(\beta +
g_{s}y_{4})]\frac{\p p}{\p y_2}  \\ 
 \nonumber
 &+& \big(F(x_R+\tan\frac{y_2}{2})- \frac{y_3}{\tau}\big)\frac{\p p}{\p y_3}\\ 
 \nonumber
   &+&  \big(F(x_R+\tan\frac{y_1}{2})- \frac{y_4}{\tau}\big)\frac{\p
     p}{\p y_4} = 0, \ 
\vspace*{-2ex}    
\end{eqnarray}  
where  $(y_1,y_2,y_3,y_4) \in {\cal A}$, and 
with boundary condition that $p=1$ on the boundary of $\cal{A}$ (since
the process is continuous). If one also adds an arbitrarily small
amount of noise for $X_3$ and $X_4$ (or considers those solutions of
(8) that arise from the limit of vanishing noise for
$X_3,X_4$),  and uses the positivity of the drift term $1-\cos\theta$
at $\theta=\pi$ where the diffusion coefficient
$\frac{\sigma^2}{2}(1+\cos\theta)^2$ vanishes, the solution of the linear elliptic partial differential equation (8) is
unique  and $\equiv 1$ so that  the process will eventually excape
from ${\cal A}$ with probability 1. 
Hence, the expected time $f({\bf y})$  of exit of the process from $\cal{A}$ satisfies ${\cal L}f =-1, \hskip .1 in  {\bf y} \in {\cal A} $ with boundary
condition $f=0$ on the boundary of $ {\cal A}$. This mean value
corresponds closely with the expected value of $T_L$ depicted in
Figure 3. In fact, for small
noise, the logarithm of the expected exit time from $ {\cal A}$, that is,
the time at which firing stops,  behaves like the inverse of the
square of the noise amplitude \cite{Fre}. When the process escapes from ${\cal A}$, it has to move into the
basin of attraction of the stable rest point. With a small probability,  noise can
eventually also drive the process out of that latter basin, so that
some intermittent spiking behavior may result. Near the bifurcation
value $g_s=g_s^\ast$, however, the situation is not symmetric between
the two attractors. The width of the basin of attraction of the stable
rest point is always positively bounded from below;
 while just beyond
the bifurcation value,  the antiphase oscillator basin of attraction
is very narrow because it emerges from two heteroclinic orbits linking
the fixed points -- the rest-points and the thresholds, and so, noise can relatively easily drive
the dynamics out of it.  Numerically we identified that the 
region of easiest escape from that basin is near the points
$P_1$ and $P_2$ in Figure 2 for the given values of the parameters. This in fact, would be where the basin is the narrowest. 
In another approach we have investigated the system of differential equations for the moments \cite{RT} of the
system (3)-(6) as given in the appendix. Numerical solutions showed that the variance of $X_1$ and $X_2$ suddenly
became extremely large in the vicinity of the exit-points $P_1$ and $P_2$ of Figure 2. 

{\it Discussion.}---
We have studied the effect of noise in systems of two coupled neurons
of Type 1.  Since the
quadratic integrate and fire neurons represent the canonical model for type I excitability, 
our results are generic for that whole class of models . We found that while coupling can support asynchronous oscillatory
activity in excitable neurons, noise can transiently terminate that sustained
spiking (near to the bifurcation point where the asynchronous periodic
orbit emerges). This circuit is a stochastic analogue of the
deterministic case previously studied by Gutkin et al. (2001) who showed that transient synchronization can terminate sustained
activity.  This two-neuron circuit is a minimal circuit model of self-sustained neural
activity. Such activity in the prefrontal cortex has been
proposed as a neural correlate of working memory \cite{Fu}. In numerical
simulations we have previously noted \cite{Gu3} that the transitions between
the two states can be quite assymetric, given that the circuit is close to the
bifurcation (i.e. the synaptic coupling is near the onset of sustained activity). 
Obviously for sufficiently weak noise the transition times are long: times for both
turning off the sustained activity and turning it on go to infinity as the strength of 
the noise goes to zero. Strong noise will produce intermitent excursions between
the two states, possibly with comparable transition times.  However, for a range 
of noise parameters, depending on the parameters of the circuitry (such as the value 
of the $\beta$ and the synaptic time constants), the time to turn off the activity is short
while the time to turn it back on (by the noise) is long. In fact  
previous simulations (see \cite{Gu3}) have lead us to believe that there is an optimal 
value of the noise to turn off the sustained activity without turning it on for any length
of simulation so  that the two transition times appear to be on separate time scales, 
and the noise effectively appears to turn off the sustained firing.
We developed a geometrical interpretation, showing that the relative size
and the geometry of the basin of attraction for the anti-phase oscillation is the key
 to this effect. Simulations  hint at a different scaling for the mean life time 
 of the sustained firing state and the silent state as a function of the noise strength. Hence we 
 would speculate that the tuning for the noise dependent destruction of the limit cycle 
stability is evocative of stochastic resonance phenomena and may be loosely interpreted
 as a delay of bifurcation by noise. Such delays have been previously noted for excitable 
 single neurons \cite{Ta2} and more recently for spatially extended systems \cite{Hu}.
 \vspace*{-7ex}
\  \\ 

 {\small 
  {\it Appendix}---
 For the system of 4 stochastic differential equations (3)-(6) we may deduce, for small noise,  the following
 differential equations for the first and second order moments, being the four means, denoted by
 $m_i, i=1,...,4$ and
 the 10 covariances $C_{ij}= {\rm Cov}[X_i,X_j]$, which includes the 4 variances, $V_i, i=1,...,4$.
$$\small{ \frac{dm_1}{dt}= m_1^2 + b  + gm_3  + V_1}, \hskip .1 in  \frac{dm_2}{dt}= m_2^2 + b + gm_4 + V_2$$
 $$\frac{dm_3}{dt}=-\frac{m_3}{\tau} + 1 + tanh(m_1-\xt) - \frac{\sinh(m_2-\xt)}{\cosh^3(m_2-\xt)}V_2$$
$$\frac{dm_4}{dt} = -\frac{m_4}{\tau} + 1 + \tanh(m_1-\xt) - \frac{\sinh(m_1-\xt)}{\cosh^3(m_1-\xt)}V_1$$
$$\frac{dV_1}{dt} = 4m_1V_1 + 2gC_{13} + \sigma^2, \hskip .1 in \frac{dV_2}{dt} = 4m_2V_2  + 2gC_{24}   + \sigma^2$$
$$\frac{dV_3}{dt} = \frac{2C_{23}}{\cosh^2(m_2-\xt)}  -\frac{2V_3}{\tau}, \hskip .1 in \frac{dV_4}{dt} =   \frac{2C_{14}} {\cosh^2(m_1-\xt)}  - \frac {2V_4}{\tau}  $$ 
$$\frac{dC_{12}}{dt} =2(m_1+m_2)C_{12}   + g(C_{14} + C_{32})$$
$$\frac{dC_{13}} {dt} =(2m_1- \frac{1}{\tau})C_{13} + gV_3+   \frac{C_{12}}{\cosh^2(m_2-\xt)}$$
$$\frac{dC_{14}}{dt} =(2m_1 -\frac{1}{\tau})C_{14}  + gC_{34} +  \frac{V_1}{\cosh^2(m_1-\xt)}$$ 
$$\frac{dC_{23}}{dt}  = (2m_2 - \frac{1}{\tau})C_{23} + gC_{34} + \frac{V_2}{\cosh^2(m_2-\xt)}$$
$$\frac{dC_{24}}{dt}  = (2m_2 - \frac{1}{\tau})C_{24} + gC_{34} + \frac{C_{21}}{\cosh^2(m_1-\xt)}$$
$$\frac{dC_{34}}{dt} = \frac{C_{24}}{\cosh^2(m_2-\xt)} + \frac{C_{31}}{\cosh^2(m_1-\xt)} - \frac {2C_{34}}{\tau}$$}
 \end{document}